\documentclass[showpacs,amssymb,preprint,preprintnumbers,nofootinbib,superscriptaddress]{revtex4-2}
\usepackage{amsmath}
\usepackage{amssymb}
\usepackage{graphicx}
\graphicspath{{Figures/}}
\usepackage{latexsym}
\usepackage{amsfonts}
\usepackage{url}
\usepackage{hyperref}
\usepackage{bm}
\usepackage{textcomp}
\usepackage{color}
\usepackage{bbm}
\usepackage{slashed}
\usepackage{caption}
\usepackage{array}
\usepackage{epstopdf}
\usepackage{subcaption}
\usepackage{placeins}
\usepackage{color}
\usepackage{cancel} 
\usepackage[normalem]{ulem}
\usepackage{tcolorbox}
\usepackage{soul}
\usepackage{commath}

\newcommand{\ba}{\begin{eqnarray}}
\newcommand{\ea}{\end{eqnarray}}

\newcommand{\myNorm}[1]{\Vert {#1} \Vert}

\begin{document}

\title{Schwarzschild-de Sitter black hole as a correlated qubit system via entropic identification}


\author{Ratchaphat Nakarachinda\footnote{Email: ratchaphat.n@rumail.ru.ac.th}}
\affiliation{Quantum and Gravity Theory Research Group, Department of Physics, \\Faculty of Science, Ramkhamhaeng University, 282 Ramkhamhaeng Road,\\ Hua mak, Bang Kapi, Bangkok 10240, Thailand}

\author{Lunchakorn Tannukij \footnote{Email: lunchakorn.ta@kmitl.ac.th}}
\affiliation{Department of Physics, School of Science, King Mongkut's Institute of Technology Ladkrabang, 1 Chalong Krung 1 Alley, Lat Krabang, Bangkok 10520, Thailand}

\author{Pitayuth Wongjun \footnote{Email: pitbaa@gmail.com}}
\affiliation{The Institute for Fundamental Study (IF), Naresuan University,\\ 99 Moo 9, Tah Poe, Mueang Phitsanulok, Phitsanulok, 65000, Thailand}

\author{Tanapat Deesuwan \footnote{Email: tanapat.dee@kmutt.ac.th}}
\affiliation{Quantum Computing and Information Research Centre (QX), Faculty of Science,\\ King Mongkut's University of Technology Thonburi, 126 Pracha Uthit Road,\\ Bang Mod, Thung Khru, Bangkok 10140, Thailand}


\begin{abstract}

The thermodynamic behaviours of multi-horizon black holes such as a Schwarzschild-de Sitter black hole have been one of the long-standing mysteries in gravitational physics since they involve quantum natures in gravitational systems and that the search for quantum gravity has not reached its conclusion. In this work, we seeked for a possibility of realising the Schwarzschild-de Sitter black hole as a correlated qubit system, where each of the event horizon is treated as a qubit and both of them are correlated in a way that two qubits could be. By identifying the entropies of subsystems to those of qubits, we successfully constructed the reduced density matrices of the two subsystems as well as the density matrix for the Schwarzschild-de Sitter black hole, modelled as 2-correlating qubits. Moreover, our results suggested that when the gravitational effect has its role in the qubit systems, supposedly like black holes, the correlation between qubits are constrained with a lower bound more stringent than the so-called Araki-Lieb triangle inequality. 


\end{abstract}
\maketitle{}


\newpage

\section{Introduction}\label{sec: intro}

One of the biggest ongoing searches in modern physics is how to reconcile the two main pillars in physics, namely, gravitation and quantum physics. Despite it being an enormous field of study, the conclusion on unifying them into quantum gravity remains elusive. On one hand, there is general relativity (GR) that utilises the geometry of spacetime to deterministically explain gravitation around gravitating objects, ranging from gigantic black holes to satellites flying above the earth. On the other hand, there is quantum mechanics which well predicts the probabilistic behaviours of small-scaled physical systems like that of elementary particles or a system of qubits. Although they have been individually successful, it has then been a challenging task to reconcile one with the other, thanks to their distinctive natures.

GR, the well-known current theory of gravity, is a classical theory and does not incorporate quantum nature. One of the significant behaviors of solutions in GR is that there exists a particular region in spacetime, from which even light cannot escape, namely ``black hole''. Among various kinds of black hole solution in GR, the Schwarzschild-de Sitter (Sch-dS) black hole is an interesting solution since it is potentially consistent with the observational results supporting the late-time accelerated expansion of the Universe \cite{SupernovaCosmologyProject:1998vns,SupernovaSearchTeam:1998fmf,Planck:2018vyg}. One of the remarkable features of the Sch-dS black hole is that there exist two event horizons; the black hole horizon referring to the inner one and the cosmological horizon referring to the outer one. Therefore, according to the thermal behaviour of the black hole, the Sch-dS black hole can be interpreted as two thermal systems with different temperatures \cite{Choudhury:2004ph}. This corresponds to the non-equilibrium system in which the equilibrium assumption used to derive the thermal behaviour of the black hole may be violated. Moreover, they also encounter thermodynamic instabilities due to the negative heat capacity and/or positive free energy \cite{Anusonthi:2025dup}. Various investigations have been conducted to describe this characteristic. In \cite{Tannukij:2020njz}, for example, the black hole horizon subsystem was assumed to be in quasi-equilibrium and its thermodynamic behaviours are governed by non-extensive thermodynamics. Furthermore, such systems can be seen as a single effective system where various assumptions on {how entropy of the effective system is related to those of subsystems} \cite{Kastor:1992nn,Bhattacharya:2015mja,Kubiznak:2016qmn} and how the effective entropy changes due to the changes of entropies of the subsystems \cite{Nakarachinda:2021jxd}. Moreover, 
one may assume that the multi-horizon black holes can be in hydrostatic equilibrium if there exists gas which obeys Tsallis statistics in between the horizons (or subsystems) \cite{Chunaksorn:2024gwo}. One of the remarkable ideas to explore this characteristic of the multi-horizon black hole is that the horizon thermodynamic systems are determined by the system with a correlation via the additional terms in the entropy composition rule \cite{Volovik:2021upi,Volovik:2021iim,Singha:2021dxe,Hashemi:2021blr,Shankaranarayanan:2003ya}. Even though these investigations are performed using semi-classical techniques to determine the quantum effect in a gravitational background, the nature of the entropy and the correlations between horizons remain largely unconstrained from an information-theoretic perspective. This naturally raises the question of how horizon entropies should be combined once the underlying system is treated as a correlated quantum system.

Quantum information theory (QIT) provides a unifying framework in which physical systems are characterised in terms of states, entropy, and correlations, with information treated as a fundamental physical quantity. In this framework, the natural carriers of information are quantum systems, whose statistical properties are encoded in density matrices and quantified by entropy measures defined within quantum theory, most commonly the von Neumann entropy. Beyond its central role in the development of quantum technologies, such as quantum computation, quantum cryptography, and quantum networks, QIT has proven to be a powerful conceptual and quantitative tool in fundamental physics. In particular, it offers a precise language for analysing correlations between subsystems, including classical correlations and genuinely quantum correlations, which are captured by entropic measures such as quantum mutual information. These correlations are constrained by universal entropy inequalities, most notably the Araki–Lieb triangle inequality \cite{Araki:1970ba}, which bounds the possible joint entropy of composite quantum systems beyond those allowed by classical theory. In gravitational settings, this information-theoretic viewpoint becomes unavoidable once horizons restrict access to parts of the quantum state, forcing a description in terms of reduced density matrices and mixed states.


This situation arises concretely in black hole physics. Owing to the groundbreaking contributions of Stephen Hawking, it was discovered that when quantum field effects are considered in curved spacetime, black holes emit thermal radiation, now known as Hawking radiation [18, 19]. Hawking further argued that this process causes an initially pure state to evolve into a mixed state characterised by thermal radiation. This apparent violation of unitarity in quantum mechanics leads to the black hole information loss paradox [20], which has motivated extensive investigations over the past decades [21–28], although a comprehensive resolution remains elusive [29, 30].

In this work, we investigate the possibility of determining the correlation of two thermal systems of the Sch-dS black hole by interpreting the system as interacting qubits. In particular, the entropy of each system is modelled as the von Neumann entropy with a specific form of the density matrix. Then, it is possible to find the bounds of the correlation of such qubits associated with the Araki-Lieb triangle inequality. With a requirement of positive semi-definiteness of the total density matrix, we found a stronger lower bound of the correlation compared to the lower bound of the Araki-Lieb triangle inequality. This suggests that the bounds of the Araki-Lieb triangle inequality must be modified due to the effects of gravity, except for the extremal case. We also found that it is possible to obtain a stronger quantum correlation in the larger black hole (closer to the extremal black hole), while it will be more suppressed for a small black hole.

This paper is organised as follows.
Section~\ref{sec: basicSchdS} is devoted to a review of the thermodynamics of the Sch-dS black hole in five dimensions. 
The thermal systems defined from this black hole are then treated as correlated qubit systems in Section~\ref{sec:5D}.
The four-dimensional case is studied in Section~\ref{sec: 4D}, following the same strategy as was done for the five-dimensional case.
A discussion of some remarkable aspects of black hole thermodynamics is addressed in Section~\ref{sec: disc}. 
A conclusion to our results is presented in Section~\ref{sec: conclu}.



\section{5D Schwarzschild-\lowercase{de} Sitter black hole and its basic thermodynamics}\label{sec: basicSchdS}

The Schwarzschild-de Sitter (Sch-dS) black hole in five dimensions can be understood through the following line element \cite{Balasubramanian:2001nb,Cai:2001sn,Myung:2007my}, in which and henceforth the natural unit, $G=c=1$ is assumed.
\begin{align}
ds^2=-F(r)dt^2+\frac{1}{F(r)}dr^2 + r^2d\Omega^2_3,
\end{align}
where $d\Omega_3^2$ is a line element in a unit 3-sphere and the metric function $F(r)$ is defined as follows.
\begin{align}
F(r) = 1-\frac{m}{r^2}-\frac{\Lambda r^2}{6}=1-\frac{r^2_s}{r^2}-\frac{r^2}{r^2_c}, \label{5DSch-dS:metricf}
\end{align}
where $m$ is the reduced mass of a black hole. 
Note that the reduced mass $m$ and the cosmological constant $\Lambda$ are parameterised with their corresponding length scales $r_s\equiv \sqrt{m}$ and $r_c \equiv \sqrt{6/\Lambda}$, respectively.

\begin{figure}[h!]
\includegraphics[scale=0.5]{ 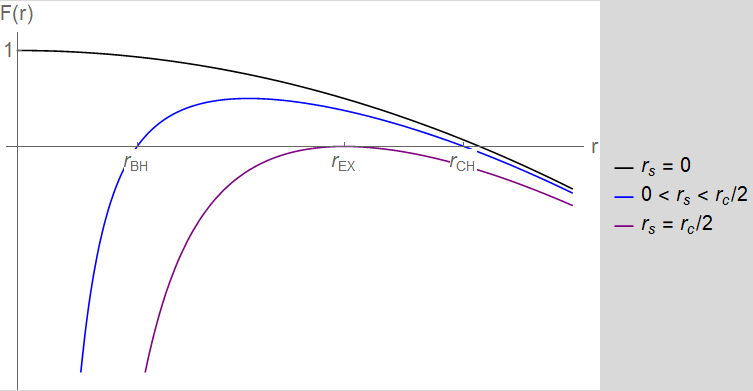}
\caption{The horizon structures of the Sch-dS metric function $F(r)$ for several cases of $r_s$ : $r_s=0$, $0<r_s<r_c/2$, $r_s = r_c/2$.}\label{Fig:horizonstructure}
\end{figure}

There are two positive zeroes for \eqref{5DSch-dS:metricf}, which represent the event horizon and the cosmological horizon located at the following radial coordinates.
\begin{align}
r_{BH} = \frac{\sqrt{r_c\left(r_c-\sqrt{r_c^2-4r^2_s}\right)}}{\sqrt{2}},\quad r_{CH} = \frac{\sqrt{r_c\left(r_c+\sqrt{r_c^2-4r^2_s}\right)}}{\sqrt{2}}, \label{Sch-dS:horizons}
\end{align}
where and henceforth the subscripts $BH,CH$ denote the corresponding quantities evaluated at the (inner) black hole horizon and the (outer) cosmological horizon, respectively. It is worth mentioning that there are two distinct event horizons for a generic value of $r_s$, given that $r_c$ is predetermined by the cosmological constant. As $r_s$ decreases to zero, so does $r_{BH}$ while $r_{CH}$ is maximised to be $r_c$. On the other hand, as $r_{s}$ increases up to its extremal limit, $r_s = r_c/2$, both event horizons merge into one and lie at $r_{EX}=\sqrt{2}\,r_s = r_c/\sqrt{2}$. 
The horizon structure for various choices of $r_s$ is shown in Fig. \ref{Fig:horizonstructure}

When treated as a thermal system, each of the event horizons of the Sch-dS black hole exhibits Hawking radiation \cite{Hawking:1975vcx}, where its thermodynamic quantities can be realised. For multi-horizon black hole systems, thermodynamic quantities are defined according to the size of the horizons. The Hawking temperatures of each horizon are proportional to the surface gravity of that particular horizon stated below \cite{Cai:2001sn,Cai:2001tv,Myung:2007my}.
\begin{align}
T_{BH/CH} = \pm\frac{1}{2\pi r_{BH/CH}}\left(1-\frac{2r^2_{BH/CH}}{r_c^2}\right),
\end{align}
where the positive and negative signs correspond to the temperature of the black hole horizon and the cosmological horizon, respectively. Moreover, the mass and entropy corresponding to each horizon can be found as \cite{Balasubramanian:2001nb,Cai:2001sn,Cai:2001tv,Myung:2007my}
\begin{align}
m_{BH/CH} = \frac{3}{8}\pi\left(r^2_{BH/CH}-\frac{r^4_{BH/CH}}{r^2_c}-\frac{r^2_{EX}}{2}\right),\qquad S_{BH/CH} = \frac{A_{BH/CH}}{4} = \frac{\pi^2 r_{BH/CH}^3}{2}.\label{Sch-dS:massandentropy}
\end{align}
Note that the entropies are proportional to the area of a 3-sphere, $A_{BH/CH}$, of radius $r_{BH/CH}$ respectively, the entropy $S_{BH/CH}$ are written in natural unit where $k_B=1$ is also assumed, and, consequently, $r_{BH}$ and $r_{CH}$ are in the unit of Planck length.

\section{The 5D S\lowercase{ch-d}S system as correlated qubits} \label{sec:5D}

In this section, the Sch-dS system is assumed to be a system of correlated qubits, where the black hole horizon subsystem is represented by one qubit and the cosmological horizon subsystem is represented by another qubit. Then, as a 2-qubit system, it is possible to find its density matrix through an analysis on the reduced density matrices of its subsystems, and determine how maximally correlated the two subsystems can be if they are treated as a qubit system.

\subsection{The Araki-Lieb triangle inequality and correlated Sch-dS system} \label{Sec:TriInEq}

{
In 1970, Araki and Lieb  \cite{Araki:1970ba} showed that the von Neumann entropies of subsystems are related to the joint von Neumann entropy of the combined system according to the entropic triangle inequality. Given a quantum-mechanical system whose density matrix is $\rho$, the von Neumann entropy $S(\rho)$ of such a system is given by
\begin{align}
S(\rho) = -\text{Tr}\left(\rho \log\rho\right),\label{vonNeumannS}
\end{align}
where $\log$ in this context is the matrix logarithm, the inverse of the matrix exponential. 
Given that $\rho_A,\rho_B$ are reduced density matrices representing subsystems of the system given by $\rho$, $S_A\equiv S(\rho_A)$, $S_B\equiv S(\rho_B)$, and $S_{AB} \equiv S(\rho)$ the Araki-Lieb triangle inequality  reads \cite{Araki:1970ba}
\begin{align}
\left|S_A - S_B\right| \leqslant S_{AB} \leqslant S_A+S_B. \label{5DSch-dS:TriInEq}
\end{align}
If there is no correlation between the subsystems, then the second equality holds; $S_{AB}=S_A+S_B$. Similarly, we may interpret the first equality as when the subsystems maximally correlate, which causes the total entropy to reach its lower bound; $S_{AB} = \left|S_A-S_B\right|$. Moreover, the natures of how the subsystems correlate according to the triangle inequality \eqref{5DSch-dS:TriInEq} can be elaborated if one separates \eqref{5DSch-dS:TriInEq} into two following inequalities.
\begin{align}
\text{max}\left(S_A,S_B\right) \leqslant &\,\,S_{AB} \leqslant S_A+S_B, \label{5DSch-dS:TriInEqmax}
\\
\left|S_A - S_B\right| \leqslant &\,\,S_{AB}\leqslant\text{max}\left(S_A,S_B\right). \label{5DSch-dS:TriInEqmin}
\end{align}
Equation~\eqref{5DSch-dS:TriInEqmax} governs the allowed total entropy of the combined system if the corresponding subsystems are classically correlated, as \eqref{5DSch-dS:TriInEqmax} is nothing but the subadditivity condition and minimum bound of $S_{AB}$ if $S_{AB}$ were defined as the Shannon entropy \cite{shannon} (see also \cite{nielsen00}). This means that if the two subsystems are classically correlated, they can correlate as much as their joint entropy will be the entropy of the ``fuzzier'' subsystem. That leaves the other inequality \eqref{5DSch-dS:TriInEqmin} to fall into quantum correlation category. As automatically suggested, the bound in \eqref{5DSch-dS:TriInEqmin} suggests a quantum correlation between subsystems and the lower bound, $\left|S_A-S_B\right|$, then indicates the joint entropy of the maximal quantum correlation between subsystems \cite{Araki:1970ba} (see also \cite{nielsen00}). This means that if the two subsystems are allowed to be quantum-correlated, then the joint entropy may reach its minimum $\left|S_A-S_B\right|$. Note that the previous statement is basically a quantum analogue of the minimum condition of the classical Shannon entropy.}

Let us assume that the Sch-dS system is a correlated system, either classically or quantum-mechanically, consisting of a 5D  black hole subsystem characterised by its event horizon $r_{BH}$ and a 5D cosmological horizon subsystem defined by its cosmological horizon $r_{CH}$. Hence, the total entropy of the 5D Sch-dS system, $S_{Total}$, must satisfy the inequality in \eqref{5DSch-dS:TriInEq} as
\begin{align}
S_{CH}-S_{BH}\leqslant S_{Total}\leqslant S_{CH}+S_{BH}, \label{5DSch-dS:Sconstraint}
\end{align}
where the lower bound is justified accordingly since $r_{CH}\geqslant r_{BH}$, and from \eqref{Sch-dS:massandentropy}, $S_{CH}\geqslant S_{BH}$.  Thus, let us parameterise $S_{Total}$ by introducing the correlation ``degree'', $a$, as follows.
\begin{align}
S_{Total}=  S_{CH}+ a S_{BH}. \label{5DSch-dS:Sa}
\end{align}
Given that $r_{CH}\geqslant r_{BH}$ and thus $S_{CH}\geqslant S_{BH}$, the parameter $a$ ranges from $-1$ to $1$ where $a=-1$ denotes the quantum maximum correlation, $a=0$ indicates the classical maximum correlation, and $a=1$ simply leads to a system of zero correlation. Note that $a$, in general, can be a function of the parameters of the system, {like $r_s$ and $r_c$ in our case.}

\subsection{Modelling Sch-dS subsystems as qubits} \label{sec:subsystemsasqubits}

If the Sch-dS system is to be treated as a quantum system of interacting qubits, then it should be represented by a density matrix $\rho_{Total}$, whose entropy is $S(\rho_{Total})$. 
Since the Sch-dS system consists of two subsystems representing the subsystem at the black hole horizon, $r_{BH}$, and that at the cosmological horizon, $r_{CH}$, then the density matrix, $\rho_{Total}$, of the combined system should contain information on the quantum states of the subsystems. In particular, one should be able to find the reduced density matrices of each of the subsystems by tracing out all the possible microstates of another subsystem.
\begin{align}
\rho_{BH} = \text{Tr}_{CH}\; \rho_{Total}, \qquad \rho_{CH} = \text{Tr}_{BH} \;\rho_{Total}\label{DenMat:partialtrace}
\end{align}
where the subscripts underneath the trace operation denotes that the partial trace is done on that particular sector. Generally speaking, if the $BH$ and $CH$ subsystems are modelled by an ${M}$-level qudit  and ${N}$-level qudit  respectively, the corresponding reduced density matrices have dimensions of ${M}\times {M}$ and ${N}\times {N}$, resulting in the dimension of $\rho_{Total}$ being ${MN}\times {MN}$.

\subsubsection{Determining $M$ and $N$}

According to \eqref{Sch-dS:massandentropy} and \eqref{5DSch-dS:Sa}, it can be seen that the entropy of a (correlated) Sch-dS system is as follows.
\begin{align}
S_{Total}&=S_{CH}+a S_{BH}, \nonumber
\\
&=\frac{\pi^2}{2}\left[\left(\frac{\sqrt{r_c\left(r_c+\sqrt{r_c^2-4r^2_s}\right)}}{\sqrt{2}}\right)^3+a\left(\frac{\sqrt{r_c\left(r_c-\sqrt{r_c^2-4r^2_s}\right)}}{\sqrt{2}}\right)^3\,\right], \nonumber
\\
&=\frac{\pi^2 r^3_c}{4\sqrt{2}}\left[\left( 1+\epsilon\sqrt{2-\epsilon^2} \right)^{3/2}+a(\epsilon) \left( 1-\epsilon\sqrt{2-\epsilon^2} \right)^{3/2}\right], \label{5DSch-dS:Sminsub}
\end{align}
where we defined a dimensionless parameter $\epsilon$ so that $r_s = \left(1-\epsilon^2\right)r_c/2$. Note that since $r_s$ should satisfy $0\leqslant r_s \leqslant r_c/2$, the dimensionless parameter $\epsilon$ must obey $0\leqslant\epsilon\leqslant1$ and the extremal limit $r_s=r_c/2$ is reached when $\epsilon=0$. Note also that in this case, we consider the correlation degree $a$ to be a function of $\epsilon$, i.e. $a=a(\epsilon)$.

\begin{figure}[h!]
\includegraphics[scale=0.5]{ 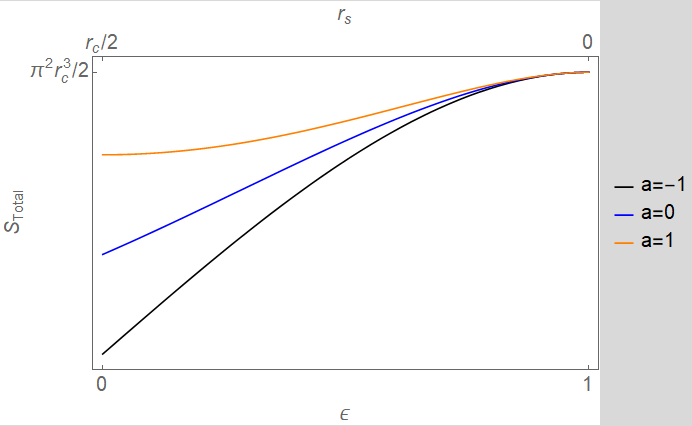}
\caption{The total entropy of the correlated Sch-dS system as a function of the dimensionless variable $\epsilon$ at different values of ``constant'' correlation degree $a$.}\label{Fig:Stotalminprofile}
\end{figure}

One of the interesting properties of $S_{Total}$ from \eqref{5DSch-dS:Sminsub} is that when $\epsilon = 1$, given that $r_c$ is a constant as defined from the cosmological constant, not only $S_{Total}$ achieves its global maximum (for the allowed range of $a$), but $S_{CH}$ is also maximised while $S_{BH}=0$ as visualised as examples in Fig. \ref{Fig:Stotalminprofile} when $a$'s are constant.
When $0\leqslant\epsilon<1$, one can view this situation as a subsystem of Schwarzschild black hole correlated with another subsystem defined by the cosmological constant. On the other hand, when $\epsilon=1$, the entropy of the system is that of the subsystem $CH$. In other words, in a universe without any black hole,
the cosmological horizon is pushed the farthest, achieving maximum entropy of $\pi^2 r^3_c/2$. If one would like to model the Sch-dS system as a qudit system whose density matrix behaves according to the behaviors mentioned above, one may model the $CH$ subsystem when it reaches its state of maximum entropy to be a maximally-mixed qudit with its corresponding reduced density matrix. The reduced density matrix must be a function of $r_s$ in such a way that the corresponding entropy of $CH$ subsystem is maximised when $r_s=0$ $(\epsilon=1)$. Thus, it is tempting to model the $CH$ subsystem with a diagonal reduced density matrix whose the diagonal elements are equally distributed, as in \eqref{DenMat:CH:max}.
\begin{align}
    \rho_{CH}(r_s=0) = \frac{1}{N} \mathbb{I}_N,
\label{DenMat:CH:max}
\end{align}
where $N$ is the number of available eigenstates of the corresponding qudit, and $\mathbb{I}_N$ is an identity matrix of dimension $N$. 


 From \eqref{vonNeumannS} and \eqref{DenMat:CH:max}, the von Neumann entropy of the reduced density matrix of the $CH$ subsystem when $r_s=0$ $(\epsilon=1)$ is 
\begin{align}
S(\rho_{CH}(r_s=0)) = \ln N.
\end{align}
On the other hand, when $r_s=0$ $(\epsilon=1)$, $S_{CH}=\pi^2 r^2_c/2$. Thus, if $CH$ subsystem is to be modelled by the reduced density matrix, $S_{CH}$ should be identified with $S(\rho_{CH})$, yielding the appropriate value of $N$,
\begin{align}
\frac{\pi^2 r^3_c}{2}&=\ln N,
\\
N &= e^{\pi^2 r^3_c/2}.
\end{align}
One of the important consequences of such an identification is that the number of eigenstates, $N$, of the qudit that represents the $CH$ subsystem is determined by the length scale $r_c$ of the cosmological constant. The other is, consequently, that since the entropy of the $BH$ subsystem is always less than that of $CH$ subsystem for generic values of parameters, $r_s,r_c$, and the two subsystems share the same value of entropy at the extremal limit, then the qudit that represents the $BH$ subsystem may have its number of eigenstate not greater than that of the qudit for the $CH$ subsystem, i.e. their reduced density matrices share the same dimension, ${M}={N}$.\footnote{Due to their entropic profiles, the $BH$ subsystem may be modelled by a qudit of fewer number of eigenstate, hence fewer number of dimension of its reduced density matrix. However, we can always extend the number of dimensions of the reduced density matrix up to that of the $CH$ subsystem by adding zero eigenvalues to the $BH$ reduced density matrix.}

Moreover, if one were to construct a complete density matrix for the Sch-dS system based on the proposed reduced density matrices, the resulting density matrix would have dimensions of ${N}^2$. Given that the length scale of the cosmological constant is well-known to be significantly large, computing for an ${N}^2\times {N}^2$ density matrix would be nearly impossible. For the sake of exploring the possibility of constructing the desired density matrix for the Sch-dS system, we greatly simplified the calculation by henceforth considering a toy model of the Sch-dS system in which $r_c = \left(\frac{2}{\pi^2}\ln 2\right)^{1/3}$ so that once it is modelled as a 2-qubit system, the number of eigenstates of each qubit is ${N}=2$.

\subsubsection{The reduced density matrices for $CH$ and $BH$ subsystems}

As mentioned in the previous subsection, we chose the situation where $r_c=\left(\frac{2}{\pi^2}\ln 2\right)^{1/3}$, so that ${N}=2$, i.e. the number of levels of the qubit in our toy model. In this subsection, since solving for the reduced density matrices of the $CH$ and $BH$ subsystems can be complex due to how complicated the von Neumann entropy is, the calculation for the density matrices will be computed numerically.

To compute the reduced density matrices, one may consider identifying the von Neumann entropy of an assumed reduced density matrix with the entropy of the corresponding subsystem. As for the $CH$ subsystem, the entropy  is given by
\begin{align}
S_{CH} &= \frac{\pi^2}{2}\left(\frac{\sqrt{r_c\left(r_c+\sqrt{r_c^2-4r^2_s}\right)}}{\sqrt{2}}\right)^3,
\\
&=\frac{\pi^2r_c^3}{2}\frac{\left(1+\epsilon\sqrt{2-\epsilon^2}\right)^{3/2}}{2\sqrt 2}, \label{SCHepsilon}
\end{align}
where the dimensionless variable $\epsilon$ which satisfies $r_s = \left(1-\epsilon^2\right)r_c/2$ has been used in the second equality. 

Recalling that the diagonal reduced density matrix of the form in \eqref{DenMat:CH:max} gives the maximum entropy, this suggests that adding non-zero non-diagonal elements to the density matrix should only result in a smaller value of entropy. Thus, the reduced density matrix for the $CH$ subsystem was chosen to follow the ansatz below.
\begin{align}
\bar\rho_{CH}=\left(
\begin{array}{cc}
\frac{1}{2}\;\;&f(\epsilon)
\\
f(\epsilon)\;\;&\frac{1}{2}
\end{array}
\right),\label{DenMat:CH:ansatz0}
\end{align}
where $f(\epsilon)$ is a function to be determined. 
The reduced density matrix of the form in \eqref{DenMat:CH:ansatz0} can always be diagonalised by rewriting it in its eigenbases, resulting in the following diagonalised reduced density matrix. 
\begin{align}
\rho_{CH} =  \left(
\begin{array}{cc}
\frac{1}{2}-f(\epsilon)&0
\\
0&\frac{1}{2}+f(\epsilon)
\end{array}
\right).\label{DenMat:CH:ansatz}
\end{align}

\begin{figure}
\includegraphics[scale=0.5]{ 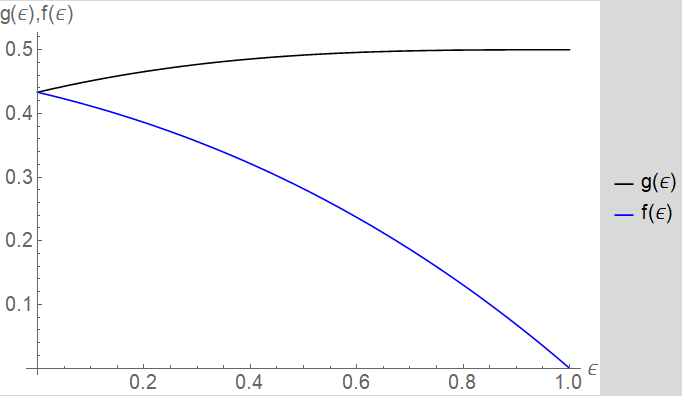}
\caption{The numerical solutions for $f(\epsilon)$, represented by the blue line, and $g(\epsilon)$, represented by the black line, at $101$ discrete values of $\epsilon$ within the range $[0,1]$.}\label{Fig:fgComparison5D}
\end{figure}

Thus, with the appropriate choice of $f(\epsilon)$, one can identify the entropy of the $CH$ subsystem and the von Neumann entropy corresponding to $\rho_{CH}$ as follows.
\begin{align}
S(\rho_{CH}) = S_{CH}. \label{DenMat:CH:Scompare}
\end{align}
According to \eqref{vonNeumannS}, \eqref{SCHepsilon} and \eqref{DenMat:CH:ansatz}, \eqref{DenMat:CH:Scompare} results in a transcendental equation. Consequently,  \eqref{DenMat:CH:Scompare} was solved numerically in a given discretised range of $\epsilon$. To this end, the allowed range of $\epsilon \in [0,1]$ is discretised into $100$ equally-spaced intervals, giving $101$ discrete values of $\epsilon$ (including $0$ and $1$). Then, \eqref{DenMat:CH:Scompare} was subsequently solved independently at each of these $101$ points, yielding a point-by-point solution over the entire domain. As a result, one can solve for $f(\epsilon)$ accordingly as the blue curve shown in Fig.~\ref{Fig:fgComparison5D}.

Similarly to the $CH$ subsystem, the same procedure can be used to find the corresponding reduced density matrix for the $BH$ subsystem. Recall that the entropy of the $BH$ subsystem is
\begin{align}
S_{BH} &= \frac{\pi^2}{2}\left(\frac{\sqrt{r_c\left(r_c-\sqrt{r_c^2-4r^2_s}\right)}}{\sqrt{2}}\right)^3,
\\
&=\frac{\pi^2r_c^3}{2}\frac{\left(1-\epsilon\sqrt{2-\epsilon^2}\right)^{3/2}}{2\sqrt 2}. \label{SBHepsilon}
\end{align}
We required that the $BH$ subsystem be modelled by a 2-level qubit system whose reduced density matrix is of the following ansatz.
\begin{align}
\rho_{BH} = \left(
\begin{array}{cc}
\frac{1}{2}-g(\epsilon)&0
\\
0&\frac{1}{2}+g(\epsilon)
\end{array}
\right), \label{DenMat:BH:ansatz}
\end{align}
where $g(\epsilon)$ is a solution to
\begin{align}
S(\rho_{BH}) = S_{BH}. \label{DenMat:BH:Scompare}
\end{align}
By solving \eqref{DenMat:BH:Scompare} numerically in a discretised range of $\epsilon$ as outlined previously, we obtained a numerical solution for $g(\epsilon)$ as the black curve shown in Fig.~\ref{Fig:fgComparison5D}.

It can be seen from Fig.~\ref{Fig:fgComparison5D} that $f(\epsilon)$ monotonically decreases as $\epsilon$ increases and reaches $0$ when $\epsilon=1$ while $g(\epsilon)$ monotonically increases as $\epsilon$ increases and then converges to $1/2$ as $\epsilon=1$. According to our assumption, this means that at $\epsilon=1$ the $CH$ subsystem behaves like an ensemble of qubits in which the quantum states of half of them are in one eigenstate and those of the other half are in another eigenstate. On the other hand, the $BH$ subsystem behaves as a pure state at $\epsilon=1$. When $\epsilon=0$, both $f(\epsilon)$ and $g(\epsilon)$ are equal to $0.433282$ at $\epsilon=0$, which denotes the extremal limit where the two horizons coincide. Alternatively, both of the subsystems can be viewed as similar ensembles of qubits, which are mixed states. Note also that $g(\epsilon)$ is always greater than $f(\epsilon)$ for the entire range of $\epsilon$, which will affect the positivity condition in Section~\ref{DenMat:construct}.

\subsection{Sch-dS system as a single interacting two-qubit system} \label{sec:systemsastwoqubits}
We have modelled both of the subsystems as qubit systems, and their reduced density matrices have been found. As a combined system, the density matrix, $\rho_{Total}$, must be a $4\times4$ square matrix. Knowing the density matrix allows one to find its corresponding reduced density matrices of the subsystems straightforwardly. Conversely, however, knowing both reduced density matrices, namely $\rho_{BH}$ and $\rho_{CH}$ may lead to various forms of density matrix of the combined system, especially since the subsystems can be correlated in numerous ways as long as the entropy of the combined system satisfies the Araki-Lieb triangle inequality in \eqref{5DSch-dS:TriInEq}. Thankfully, \cite{PhysRevA.93.062320} showed that there is a systematic way to deal with such freedom through non-local transformations on the density matrix which introduces correlation between the two subsystems.
Henceforth, we will follow the formulation of the density matrix for a system of two interacting qubits found in \cite{PhysRevA.93.062320}.

\subsubsection{Density matrix in Dirac basis}

A $4\times 4$ density matrix of two interacting qubits system can be generally written in the Dirac basis, whose elements, dubbed Dirac matrices, $D_{\mu\nu}$, are defined as 
\begin{align}
D_{\mu\nu} = \sigma_\mu \otimes \sigma_{\nu},\label{DenMat:Bases}
\end{align}
where $\sigma_\mu \in \left(\sigma_0,\sigma_1,\sigma_2,\sigma_3\right)$ as $\mu$ ranges from $0,1,2,3$ and $\sigma_\mu$'s are extended Pauli matrices, which are three Pauli matrices and an identity matrix, as follows.
\begin{align}
\sigma_0 =I=\begin{pmatrix}
\;1\;&\;0\;
\\
\;0\;&\;1\;
\end{pmatrix},\qquad \sigma_1 =\begin{pmatrix}
\;0\;&\;1\;
\\
\;1\;&\;0\;
\end{pmatrix},\qquad \sigma_2 =\begin{pmatrix}
\;0\;&-i
\\
\;i\;&\;0\;
\end{pmatrix},\qquad \sigma_3 =\begin{pmatrix}
\;1\;&\;0\;
\\
\;0\;&-1\;
\end{pmatrix}.
\end{align}
Consequently, the Dirac basis consists of $16$ elements which satisfy the following orthogonal relation:
\begin{align}
\text{Tr}\left(D_{\alpha\beta}D_{\gamma\delta}\right) = 4 \delta_{\alpha\gamma}\delta_{\beta\delta}.
\end{align}
In this basis, the $4\times 4$ density matrix, $\rho_{Total}$, can be written as
\begin{align}
\rho_{Total} = \frac{1}{4} r_{\mu\nu} D_{\mu\nu},
\end{align}
where the repeated indices are assumed to be summed over possible values, and the $16$ scalar coefficients, $r_{\mu\nu}$, form a $4\times 4$ matrix called Bloch matrix, $\overset{\leftrightarrow}{r}$.

The Bloch matrix, $\overset{\leftrightarrow}{r}$, can be split into four components, namely the scalar of unit value, two 3-dimensional vectors, and a $3\times 3$ matrix as follows.
\begin{align}
\overset{\leftrightarrow}{r} = \left[\begin{array}{c|ccc}
1&r_{01}&r_{02}&r_{03}
\\
\hline
r_{10}&r_{11}&r_{12}&r_{13}
\\
r_{20}&r_{21}&r_{22}&r_{23}
\\
r_{30}&r_{31}&r_{32}&r_{33}
\end{array}\right] \equiv 
\begin{bmatrix}
\;1\;&\;\vec{v}\,^\dagger\;
\\
\;\vec u\; & \;R\;
\end{bmatrix},
\end{align}
where $u_i = r_{i0}$, $v_j = r_{0j}$, $R_{ij}=r_{ij}$, $i,j\in\left\{1,2,3\right\}$, and $\dagger$ denotes the Hermitian conjugate. Note that $r_{00} = 1$ to ensure that the trace of the density matrix $\rho_{Total}$ is unity. $\vec u,\vec v$ represent Bloch vectors corresponding to reduced density matrices of the subsystems, while $R$ represents the correlation matrix between the two subsystems.

\subsubsection{Local and non-local unitary transformations}

Once the density matrix, $\rho_{Total}$, of a combined system is obtained, one can always change the basis of the density matrix by applying unitary transformations that belong to the $SU(4)$ group. Such transformations leave the eigenvalues of $\rho_{Total}$ invariant. The unitary transformation $U$ can be written as products of local operators and non-local operators as follows \cite{PhysRevA.93.062320,Zhang:2003zz}.
\begin{align}
U = \left(U_1\otimes U_2\right)\overset{\circ}{U}_1(\theta_1)\overset{\circ}{U}_2(\theta_2)\overset{\circ}{U}_3(\theta_3)\left(U_3\otimes U_4\right),
\end{align}
where $U_i\otimes U_j \in SU(2)\otimes SU(2)$ are local operators whose $U_i,U_j$ act on each qubit separately, while $\overset{\circ}{U}_i(\theta_i)$ are non-local operators that perform basis transformation across the two qubits, given by
\begin{align}
\overset{\circ}{U}_i(\theta_i) \equiv exp\left(\frac{i}{2}\theta_i \sigma_i \otimes \sigma_i\right),
\end{align}
where $\theta_i$ is a parameter of this transformation, and repeated indices here do not imply summation. The transformed density matrix is then
\begin{align}
\rho'_{Total} = U \rho_{Total} U^\dagger.
\end{align}
In order to grasp the nature of correlation between the two subsystems, one needs to do it through an application of non-local transformations\footnote{The local transformation only perform a change of basis locally in the subsystem, and do not affect towards correlations across subsystems.}. It can be best understood through an uncorrelated system, the so-called product state.
Given a system whose subsystems, in this context $BH$ and $CH$, are not correlated, the density matrix of the product state satisfies the followings.
\begin{align}
\rho_{Total} = \rho_{CH} \otimes \rho_{BH}, \label{5DSch-dS:ProductState}
\end{align}
If one finds the entropies of the combined system and the corresponding subsystems, one will find that the entropies satisfy the upper bound of the triangle inequality in \eqref{5DSch-dS:TriInEq}.
Performing the non-local transformation, for simplicity, $\overset{\circ}{U}_1(\theta_1)$ on $\rho_{Total}$ then introduces correlation across the $BH$ and $CH$ subsystems. In particular, the non-local transformation leaves the eigenvalues of the density matrix, as well as its entropy, invariant, while the non-local transformation introduces the correlation by altering the entropies of subsystems, which in turn alters the entropic composition rule of the system so that it satisfies the triangle inequality in \eqref{5DSch-dS:TriInEq}.

\subsubsection{The density matrix construction} \label{DenMat:construct}

Before a density matrix of the combined system can be constructed, knowledge of basis used to express the density matrix is necessary to unambiguously perform partial traces to obtain the reduced density matrices of the subsystems. We then assigned the numbers $x_i,y_j\in \{1,2\}$ to the ket state $\left|x_iy_i\right>$ (and the bra state $\left<x_iy_j\right|$) and use them as the basis, where $x$ denotes the possible state of the subsystem $CH$ and $y$ denotes that of the subsystem $BH$. Consequently, the basis on which the density matrix is appropriately expressed is as follows.
\begin{align}
&\;\;\;\;\begin{array}{wc{1cm}wc{1cm}wc{1cm}wc{1cm}}
\left<11\right|&\left<12\right|&\left<21\right|&\left<22\right|
\end{array}\nonumber
\\
\rho_{Total} = \begin{array}{wc{1cm}}
\left|11\right>
\\
\left|12\right>
\\
\left|21\right>
\\
\left|22\right>
\end{array}&\left(\begin{array}{wc{1cm}wc{1cm}wc{1cm}wc{1cm}}
&&&
\\
&&&
\\
&&&
\\
&&&
\end{array}\right).
\end{align}
Then, partial traces can be clearly defined as
\begin{align}
\left(\rho_{BH}\right)_{y_1y_2}=\sum^2_{x=1}\left<xy_1\right| \rho_{Total}\left|xy_2\right>,\quad \left(\rho_{CH}\right)_{x_1x_2}=\sum^2_{y=1}\left<x_1y\right| \rho_{Total}\left|x_2y\right>.
\end{align}

As previously suggested, we started with constructing the density matrix of a product state as shown in \eqref{5DSch-dS:ProductState} \cite{PhysRevA.93.062320}. {Here, we assumed that the functions that characterise the reduced density matrices before introducing the correlation are $\bar f$ for $CH$ subsystem, and $\bar g$ for $BH$ subsystem.} The Bloch matrix corresponding to the density matrix of the product state reads
\begin{align}
\overset{\leftrightarrow}{r} = \left[\begin{array}{wc{.8cm}|wc{.8cm}wc{.8cm}wc{.8cm}}
1&0&0& -2 \bar{g}
\\
\hline
0&0&0&0
\\
0&0&0&0
\\
-2 \bar{f}&0 &0 &4 \bar{f}\bar{g}
\end{array}\right].
\end{align}
The resulting density matrix then reads
\begin{align}
\bar\rho_{Total} &=
\begin{pmatrix}
    \frac{1}{4}\left(1-2\bar{f}\right)\left(1-2\bar{g}\right)&0&0&0
    \\
    0&\frac{1}{4}\left(1-2\bar{f}\right)\left(1+2\bar{g}\right)&0&0
    \\
    0&0&\frac{1}{4}\left(1+2\bar{f}\right)\left(1-2\bar{g}\right)&0
    \\
    0&0&0&\frac{1}{4}\left(1+2\bar{f}\right)\left(1+2\bar{g}\right)
\end{pmatrix}. \label{DenMat:Total}
\end{align}
The corresponding partial traces, yielding { ``uncorrelated" reduced density matrices}, $\bar\rho_{BH},\bar\rho_{CH}$, can be found as
\begin{align}
\bar\rho_{BH}  = \begin{pmatrix}
\frac{1}{2}-\bar{f}&0
\\
0&\frac{1}{2}+\bar{f}
\end{pmatrix},\quad \bar\rho_{CH} = \begin{pmatrix}
\frac{1}{2}-\bar{g}&0
\\
0&\frac{1}{2}+\bar{g}
\end{pmatrix}.\label{DenMat:Total1}
\end{align}
In addition, the density matrix $\bar\rho_{Total}$ represents the product state in the sense that the density matrix can be expressed as a tensor product of the reduced density matrices.
\begin{align}
\bar\rho_{Total} = \bar\rho_{CH}\otimes \bar\rho_{BH}, \label{DenMat:Total2}
\end{align}
Note that $\bar\rho_{Total}$ represents an uncorrelated system whose entropy satisfies the upper bound of Araki-Lieb triangle inequality in \eqref{5DSch-dS:TriInEq}. 
The correlation can be introduced by the application of a non-local transformation. For simplicity, we considered a non-local transformation $\overset{\circ}{U}_1(\theta_1)$ on the density matrix $\bar\rho_{Total}$ as follows\footnote{The result is the same if $\overset{\circ}{U}_2(\theta_2)$ is used instead of $\overset{\circ}{U}_1(\theta_1)$ while $\overset{\circ}{U}_3(\theta_3)$ transforms $\bar \rho_{Total}$ trivially. We left the situation where consecutive non-local transformations are used for future works.}.
\begin{gather}
\rho_{Total} = \overset{\circ}{U}_1(\theta_1)\, \bar\rho_{Total} \,\overset{\circ}{U}^\dagger_1(\theta_1),\nonumber
\\
   \hspace{-2cm}=
{\footnotesize\begin{pmatrix}
    \frac{1}{4}\left(1-2\left(\bar f+\bar g\right)\cos\theta_1+4\bar f \bar g\right)&0&0&\frac{i}{2} \left(\bar f+\bar g\right)\sin \theta_1
    \\
    0&\frac{1}{4}\left(1-2\left(\bar f-\bar g\right)\cos\theta_1-4\bar f \bar g\right)&\frac{i}{2} \left(\bar f-\bar g\right)\sin \theta_1&0
    \\
    0&-\frac{i}{2} \left(\bar f-\bar g\right)\sin \theta_1&\frac{1}{4}\left(1+2\left(\bar f-\bar g\right)\cos\theta_1-4\bar f \bar g\right)&0
    \\
    -\frac{i}{2} \left(\bar f+\bar g\right)\sin \theta_1&0&0&\frac{1}{4}\left(1+2\left(\bar f+\bar g\right)\cos\theta_1+4\bar f \bar g\right)
\end{pmatrix} }\label{DenMat:Totalrotated}
\end{gather}
The corresponding reduced density matrices then become
\begin{align}
\rho_{BH}  = \begin{pmatrix}
\frac{1}{2}-\bar{g}\cos \theta_1&0
\\
0&\frac{1}{2}+\bar{g}\cos\theta_1
\end{pmatrix},\quad \rho_{CH} = \begin{pmatrix}
\frac{1}{2}-\bar{f}\cos\theta_1&0
\\
0&\frac{1}{2}+\bar{f}\cos\theta_1
\end{pmatrix}. \label{DenMat:BHCHrotated}
\end{align}
Note that the partial traces yield different results under different non-local transformations since the parameter $\theta_1$ modifies the probabilities of microstates corresponding to the reduced density matrices. To understand how $\theta_1$ affects the correlation, let us consider the situation where $\bar g$ and $\bar f$ are fixed. The non-local transformation, with parameter $\theta_1$, leaves the entropy of the combined system invariant while it alters the entropies of the subsystems so that it modifies the entropic composition rule of the system. In particular, as $\theta_1=0$  increases from $0$ to $\pi/2$, the subsystems become more of the maximally entropic state, where the diagonal elements are equally distributed, thus corresponding to the subsystems having more entropies.

To model the Sch-dS system as a system of two interacting qubits, we require that $\rho_{BH}$ and $\rho_{CH}$ in \eqref{DenMat:BHCHrotated} are to be identified with those in \eqref{DenMat:BH:ansatz} and in \eqref{DenMat:CH:ansatz}. We then require
\begin{align}
    \bar g = g(\epsilon)\sec\theta_1,\qquad \bar f = f(\epsilon) \sec\theta_1. \label{DenMat:BHCHmatching}
\end{align}
It is important to note that once \eqref{DenMat:BHCHmatching} is satisfied, then the effect of $\theta_1$ towards the entropic properties changes drastically from what previously mentioned. In this case, $g(\epsilon)$ and $f(\epsilon)$ will be fixed to ensure the correct values of entropies of the subsystems (as suggested in \eqref{SCHepsilon} and \eqref{SBHepsilon}). In this case, as $\theta_1$ increases from $0$ to $\pi/2$, the entropies of the subsystems do not change while the combined entropy decreases, hence the correlation between two subsystems.

\subsubsection{Positivity condition and Maximally-correlated system}

To ensure that the density matrix $\rho_{Total}$ as in \eqref{DenMat:Totalrotated} is physical, it must be positive semi-definite. In the other words, all four eigenvalues of $\rho_{Total}$ must not be negative. Since $\rho_{Total}$ and $\bar\rho_{Total}$ are related only through a unitary transformation, the previous statement is equivalent to saying that all four eigenvalues of $\bar\rho_{Total}$ are not negative. According to \eqref{DenMat:Total}, \eqref{DenMat:Total1}, and \eqref{DenMat:Total2}, it is obvious that the following constraints ensure that the density matrix $\bar\rho_{Total}$, and consequently $\rho_{Total}$, are positive semi-definite.
\begin{align}
-\frac{1}{2} \leqslant \bar g \leqslant \frac{1}{2}, \qquad -\frac{1}{2} \leqslant \bar f \leqslant \frac{1}{2}, \label{DenMat:poscond0}
\end{align}
Taking \eqref{DenMat:BHCHmatching} into account, \eqref{DenMat:poscond0} can also be translated as follows.
\begin{align}
    -\frac{1}{2g(\epsilon)} \leqslant \sec\theta_1 \leqslant \frac{1}{2g(\epsilon)}, \qquad -\frac{1}{2f(\epsilon)} \leqslant \sec\theta_1 \leqslant \frac{1}{2f(\epsilon)}.
\end{align}
Moreover, one can see from Fig.~\ref{Fig:fgComparison5D} that $g(\epsilon)>f(\epsilon)$ for the entire range of $\epsilon$, meaning that the only valid positivity condition is
\begin{align}
    -\frac{1}{2g(\epsilon)} \leqslant \sec\theta_1 \leqslant \frac{1}{2g(\epsilon)}.\label{DenMat:poscond1}
\end{align}
Without loss of generality, we can restrict our study to the case where $0<\sec\theta_1\leqslant\frac{1}{2g(\epsilon)}$ because if $\sec\theta_1$ lies in the negative range, we can always redefine the basis of all density matrices so that $g(\epsilon)\to \mathcal{G}(\epsilon)=-g(\epsilon)$, and $f(\epsilon)\to \mathcal{F} (\epsilon)=-f(\epsilon)$ and all subsequent results will match what we have obtained so far. 

\begin{figure}[h!]
\includegraphics[scale=0.5]{ 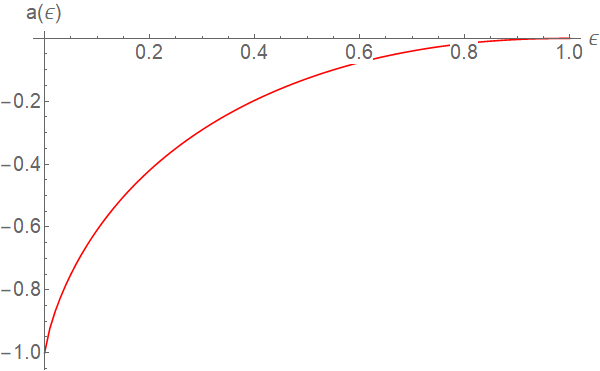}
\caption{The plot shows the behaviour of $a(\epsilon)$ at different values of $\epsilon$ in the case that the qubit system analogous to 5D Sch-dS system is maximally-correlated.}\label{Fig:aCorr5D}
\end{figure}

The positivity condition in \eqref{DenMat:poscond1} suggests how maximally-correlated the subsystems can be. 
Since $\theta_1$ introduces the correlation between the two subsystems, then $\theta_1$ determines the correlation degree $a(\epsilon)$. From the positivity condition in \eqref{DenMat:poscond1}, the two subsystem will be maximally correlated when $\theta_1$ reaches its upper bound; where $\sec\theta_1=\frac{1}{2g(\epsilon)}$ (we ignored the negative side of the positivity condition). With that value of $\theta_1$, $a(\epsilon)$ can be found as
\begin{gather}
a(\epsilon)= \frac{S(\rho_{Total})-S_{CH}}{S_{BH}},\nonumber
\\
\hspace{-1.5cm}={\small\frac{\left(1-2f(\epsilon)\right)\ln\left(1-2f(\epsilon)\right)+\left(1+2f(\epsilon)\right)\ln\left(1+2f(\epsilon)\right)-\left(1-\frac{f(\epsilon)}{g(\epsilon)}\right)\ln\left(1-\frac{f(\epsilon)}{g(\epsilon)}\right)-\left(1+\frac{f(\epsilon)}{g(\epsilon)}\right)\ln\left(1+\frac{f(\epsilon)}{g(\epsilon)}\right)}{\ln 4 -\left(1-2g(\epsilon)\right)\ln\left(1-2g(\epsilon)\right)-\left(1+2g(\epsilon)\right)\ln\left(1+2g(\epsilon)\right)}}. \label{DenMat:aCorrallowed}
\end{gather}
The behaviour of $a(\epsilon)$ at various values of $\epsilon$ is shown in Fig. \ref{Fig:aCorr5D}.
From Fig. \ref{Fig:aCorr5D}, the allowed values of $a(\epsilon)$ are determined by the maximum values of $\theta_1$, i.e. the maximum degree of correlation as allowed by the positivity condition in \eqref{DenMat:poscond1}. 
When $\epsilon=0$, the extremal limit, the entropy of the combined system can reach its lower bound as allowed by the triangle inequality, as denoted by $a(\epsilon)=-1$. However, when $\epsilon>0$, the correlation cannot be as much as at the extremal limit and the correlation degree $a(\epsilon)$ increases from $-1$ to $0$ as $\epsilon=1$ which is the configuration where there is a cosmological horizon but no black hole. Furthermore, this profile of $a(\epsilon)$ suggests that if there exists such a Sch-dS black hole in the universe, the correlation between its subsystems can be as much as indicated by \eqref{DenMat:aCorrallowed} and in Fig. \ref{Fig:aCorr5D}. 

\begin{figure}[h!]
\includegraphics[scale=0.5]{ 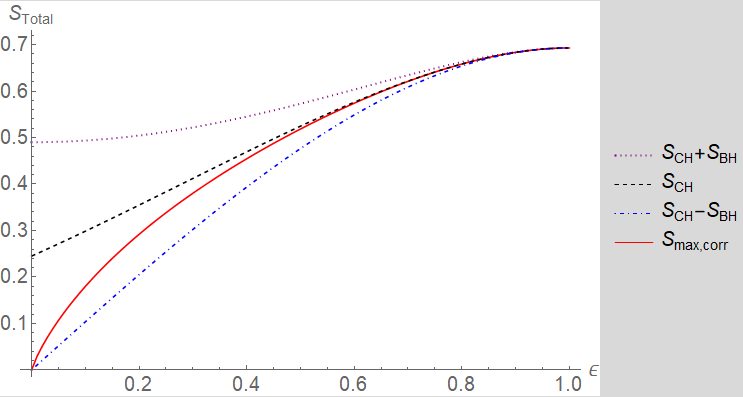} \quad \includegraphics[scale=0.5]{ 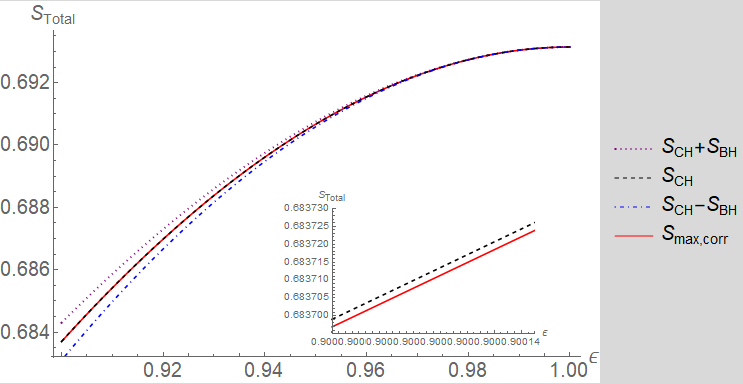}
\caption{The above figure shows the entropies in each cases of correlation between subsystems. The red line denotes the entropy of the maximally-correlated 5D Sch-dS black hole system. The below figure demonstrates a zoom-in of the above figure at  $\epsilon$ close to unity. }\label{Fig:S5Dcompare}
\end{figure}

Alternatively, the entropy corresponding to such a correlation degree is shown in Fig. \ref{Fig:S5Dcompare} among three entropic milestones as discussed in \ref{Sec:TriInEq}.
As elaborated in Fig. \ref{Fig:S5Dcompare}, if modelled as a qubit system, the subsystems of the Sch-dS black hole can be classically correlated which covers the whole classical regime in the Araki-Lieb triangle inequality in \eqref{5DSch-dS:TriInEq}. On the other hand, the subsystems may have quantum correlation within themselves for the whole range of $\epsilon$, but the total entropy cannot reach the lower bound of the Araki-Lieb triangle inequality. Moreover, if the system of two qubits is taken into account, this result suggests that the quantum correlation in qubit systems may be diminished due to the strong gravitational effect, while qubit systems without gravity can reach the lower bound of the triangle inequality without any {restriction}.

\section{The 4D S\lowercase{ch-d}S system as correlated qubits}\label{sec: 4D}

In this section, the same procedure for 5D Sch-dS black hole is employed on 4D Sch-dS black hole. The same physical  phenomena can be observed in 4D system as well, suggesting that the limitation in having quantum correlation across black hole subsystems is more or less universal, at least for 4D and 5D.

The line element in the 4D Sch-dS geometry is the following.
\begin{align}
ds^2 = -F_4(r)dt^2+\frac{1}{F_4(r)}dr^2+r^2 d\Omega^2_2, 
\end{align}
where $d\Omega^2_2$ is a line element in a unit 2-sphere and the metric function $F_4(r)$ is as follows.
\begin{align}
F_4(r) = 1-\frac{2m}{r} - \frac{\Lambda r^2}{3} \equiv 1-\frac{r_{s4}}{r} - \frac{r^2}{r_{c4}^2}, \label{4DSch-dS:metricf}
\end{align}
where we introduced the corresponding length scales, $r_{s4} \equiv 2m$ and $r_{c4}\equiv \sqrt{3/\Lambda}$, $m$ is the mass of the Sch-dS black hole in 4D, and $\Lambda$ is the cosmological constant.

\begin{figure}[h!]
\includegraphics[scale=0.5]{ 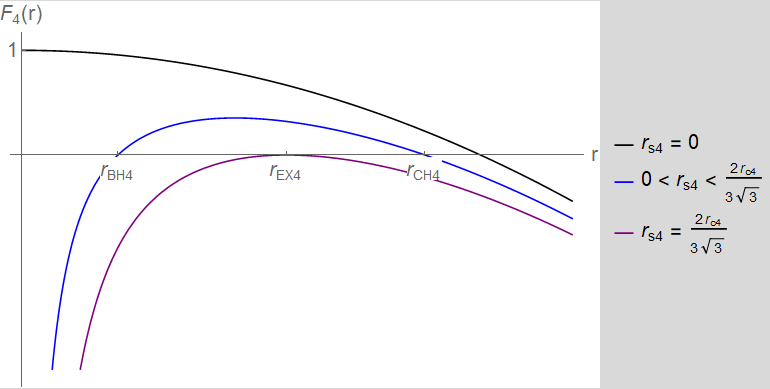} 
\caption{The horizon structures of the 4D Sch-dS metric function $F_4(r)$ for several cases of $r_s$ : $r_{s4}=0$, $0<r_{s4}<2r_{c4}/3\sqrt{3}$, $r_{s4} = 2r_{c4}/3\sqrt{3}$.}\label{Fig:horizonstructure4D}
\end{figure}

\begin{figure}[h!]
\includegraphics[scale=0.5]{ 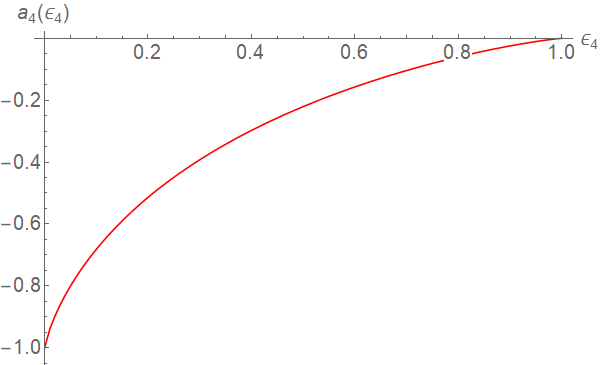}
\caption{The plot shows the behaviour of $a(\epsilon)$ at different values of $\epsilon$ in the case that the qubit system analogous to 4D Sch-dS system is maximally-correlated.}\label{Fig:aCorr4D}
\end{figure}

\begin{figure}[h!]
\includegraphics[scale=0.5]{ 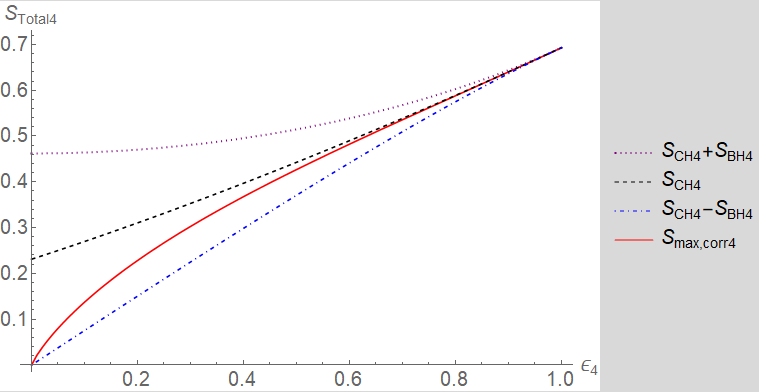} \quad \includegraphics[scale=0.5]{ 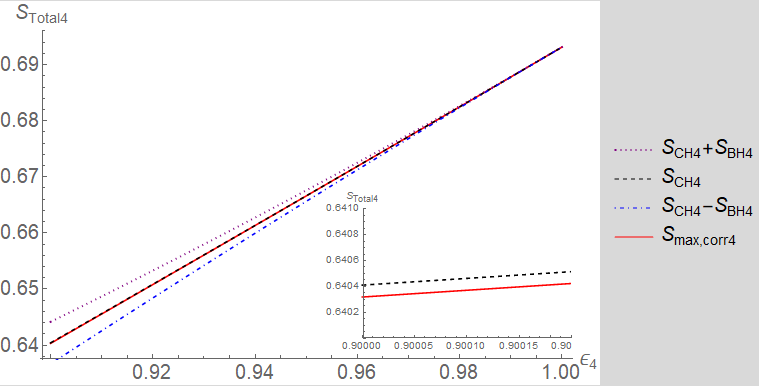}
\caption{The above figure shows the entropies in each cases of correlation between subsystems. The red line denotes the entropy of the maximally-correlated 4D Sch-dS black hole system. The below figure demonstrates a zoom-in of the above figure at  $\epsilon$ close to unity. }\label{Fig:S4Dcompare}
\end{figure}

The black hole horizon and the cosmological horizon of the 4D Sch-dS black hole correspond to the two positive zeroes to \eqref{4DSch-dS:metricf}, which basically is a depressed cubic equation. The two zeroes are as follows.
\begin{align}
r_k = 2\frac{r_{c4}}{\sqrt{3}} \cos \left[\frac{1}{3}\arccos \left(-\frac{3 \sqrt{3}r_{s4}}{2 r_{c4}}\right)-\frac{2\pi k}{3}\right],\quad k\in 0,1,2.
\end{align}
Note that the solutions {are real} when $r_{s4}\leqslant \frac{2r_{c4}}{3\sqrt{3}}$, whose upper bound is actually the condition for the extremal limit of 4D Sch-dS black hole. The black hole horizon, $r_{BH4}$, and the cosmological horizon, $r_{CH4}$, corresponding to the solution above, are as follows.
\begin{align}
r_{BH4} &= r_1 = 2\frac{r_{c4}}{\sqrt{3}} \cos \left[\frac{1}{3}\arccos \left(-\frac{3 \sqrt{3}r_{s4}}{2 r_{c4}}\right)-\frac{2\pi}{3}\right],
\\
r_{CH4} &= r_0 = 2\frac{r_{c4}}{\sqrt{3}} \cos \left[\frac{1}{3}\arccos \left(-\frac{3 \sqrt{3}r_{s4}}{2 r_{c4}}\right)\right].
\end{align}
Moreover, the 4D Sch-dS black hole also exhibits the extremal property where at its extremal limit, the two horizons merge as $r_{BH4}=r_{CH4} = 3r_{s4}/2= r_{c4}/\sqrt{3}$, as can be seen in Fig. \ref{Fig:horizonstructure4D}.
Due to the Hawking radiation, the corresponding temperature of 4D Sch-dS black hole is \cite{Gibbons:1977mu,Pappas:2017kam,Kubiznak:2016qmn,Tannukij:2020njz}
\begin{align}
T_{BH4/CH4} = \pm \frac{1}{4\pi r_{BH4/CH4}} \left(1- \frac{3r^2}{r_{c4}^2}\right).
\end{align}
Then, the corresponding masses and entropies for each horizons are \cite{Gibbons:1977mu,Pappas:2017kam,Kubiznak:2016qmn,Tannukij:2020njz}
\begin{align}
m_{BH4/CH4}=\frac{r_{BH4/CH4}}{2}\left(1-\frac{r^2_{BH4/CH4}}{r^2_{c4}}\right), \qquad S_{BH4/CH4} = \frac{A_{BH4/CH4}}{4} = \pi r^2_{BH4/CH4}. \label{4DSch-dS:massandentropy}
\end{align}
By assuming that the 4D Sch-dS system is a quantum-correlated system, then its total entropy, $S_{Total4}$, must satisfy the Araki-Lieb triangle inequality in \eqref{5DSch-dS:Sconstraint}. Thus, $S_{Total4}$ can be parameterised with the correlation degree, $a_4$, as follows.
\begin{align}
S_{Total4} = S_{CH4}+a_4 S_{BH4}, \label{4DSch-dS:entropiccomp}
\end{align}
where $S_{CH4}$ and $S_{BH4}$ are the entropies of subsystems defined in \eqref{4DSch-dS:massandentropy}. As for the corresponding correlated qubit analogue to the 4D Sch-dS system,  we can follow the same procedure mentioned in Sec.\ref{sec:subsystemsasqubits},  Sec. \ref{sec:systemsastwoqubits}, and by choosing $N=2$, the profile of the correlation degree $a_4$ corresponding to the maximally-correlated system, in terms of the dimensionless parameter $\epsilon_4$ which satisfies $r_{s4}= \frac{2r_{c4}}{3\sqrt{3}}\left(1-\epsilon^2_4\right) $, can be shown in Fig. \ref{Fig:aCorr4D}.
Moreover, the entropy of the maximally-correlated 4D Sch-dS system is shown numerically in Fig. \ref{Fig:S4Dcompare} along with both the upper bound and the lower bound of the Araki-Lieb triangle inequality, and $S_{CH4}$ which denotes the lower bound of the classically-correlated 4D Sch-dS system.
The same things, as in the case of 5D Sch-dS system, can be said in the 4D case as well. In particular, in this context, the subsystems can correlate both classically and quantum-mechanically. However, there is a more stringent lower bound to the total entropy which limits how the subsystems can have quantum correlation. The more stringent lower bounds appear both in the 5D case and in 4D case, indicating that gravitational effects alter the nature of quantum correlation between subsystems. One may speculate that since the bounds can be found in both cases, they may appear similarly in systems in other numbers of dimensions. 

\section{Discussion}\label{sec: disc}
\subsection{The Sch-dS black hole as an effective thermal system}

One of the mysteries in black hole thermodynamics is that the thermal properties of multi-horizon black holes can be ambiguous since all of the event horizons give rise to their own sets of thermodynamical quantities. Numerous attempts have been made to explain these characteristics, including the definition of an effective thermal system. In particular, one may realise the multi-horizon black holes as thermal systems whose effective thermodynamical quantities are derived from those defined on each horizon. To this end, one may assume different forms of entropy composition, which leads to different sets of effective thermodynamical quantities \cite{Kastor:1992nn,Bhattacharya:2015mja,Nakarachinda:2021jxd} (see also \cite{Kubiznak:2016qmn}). Without knowing the exact correlation between each event horizon, the entropy composition can be assumed in any form as long as it obeys the Araki-Lieb triangle inequality. According to our results, there is a more stringent bound to the entropy composition of Sch-dS black holes, leading to fewer possibilities for the entropy composition rule. Taking Fig.~\ref{Fig:aCorr5D} and Fig.~\ref{Fig:aCorr4D} into account, one can see that the entropy composition rule for a Sch-dS black hole system can be $S_{Total} = S_{CH}-S_{BH}$ only when the configuration of the system corresponds to $\epsilon=0$, or when the black holes reach their extremal limits. Away from the extremal limits, if one wants to realise the Sch-dS black holes as effective thermodynamic systems, one should assume the entropy composition rules that lie within $S_{CH}+aS_{BH}\leqslant S_{Total} \leqslant S_{CH}+S_{BH}$ where $a$ obeys the result in Fig. \ref{Fig:aCorr5D} (or Fig. \ref{Fig:aCorr4D} for $a_4$). An interesting choice that arises from our analysis is to use the entropy composition rule with a running correlation degree $a$ which corresponds to a maximally-allowed-correlated system, constrained from the positivity condition of the density matrices, and formulate the consequent effective black hole thermodynamics, which will be left for future work.

\subsection{Mutual information and non-additivity}

In this work, the entropic compositions are parameterised as in \eqref{5DSch-dS:Sa} and in \eqref{4DSch-dS:entropiccomp}. Generally speaking, the entropic composition rule of two correlated subsystems can be written as
\begin{align}
    S_{Total}=S_{CH}+S_{BH}-S_{corr}(S_{CH},S_{BH}),
\end{align}
where $S_{corr}$ represents mutual information of the system that is a function of both entropies of each subsystem, that is, $S_{CH}$ and $S_{BH}$. If $S_{corr}$ is assumed to be of the product form, namely, $S_{corr}\sim S_{CH}S_{BH}$, then the total entropy, parameterised with $a$, can be expressed as
\begin{align}
    S_{Total}
    &=S_{CH}+aS_{BH}
    =S_{CH}+S_{BH}-\frac{(1-a)}{S_{CH}}S_{BH}S_{CH},\nonumber\\
    &\equiv S_{CH}+S_{BH}-b(\epsilon)S_{BH}S_{CH}.
\end{align}
With the results in Fig. \ref{Fig:aCorr5D} and in Fig. \ref{Fig:aCorr4D}, one may evaluate how dependent one subsystem is to the other, the so-called mutual information, of the Sch-dS black hole system numerically. 

Moreover, one may be able to interpret the correlation degree, $a$, as a deviation from the entropic additivity in the context of Tsallis entropy \cite{Tsallis:1987eu}. Taking into account the Tsallis composition rule,
\begin{align}
    S_\lambda(A,B) = S_\lambda (A) + S_\lambda(B) + \lambda S_\lambda (A) S_\lambda (B),
\end{align}
where $\lambda$ measures the departure from additivity, or the "non-additivity", of the system. One may translate the correlation degree, $a$, in terms of non-additivity as
\begin{align}
    \lambda=-b(\epsilon) = \frac{a-1}{S_{CH}}.
\end{align}
One important point of this translation that needs further discussion is that the mutual information is not truly a function of both the entropies of the subsystems. In order to have more accurate translation, one needs to broaden the scope of this work to cover the case of cosmological horizon with arbitrary size, which will be left as a future work.


\section{Conclusion}\label{sec: conclu} 

We have investigated the possibility of realising a composition of thermal systems representing a black hole with two horizons, especially the Sch-dS black hole. Both of the event horizons of Sch-dS black hole are considered to be two thermal subsystems. Each subsystem is then modelled as a qudit by assuming that the qudit system and the gravitational subsystem share the same entropy. We required that when the subsystem defined by the cosmological horizon has maximum entropy, the corresponding reduced density matrix must be the one with maximum entropy, i.e. the one whose diagonal elements are equally distributed, which determines the number of levels of the corresponding qudit, as well as the other qudit for the event horizon subsystem. We then considered a toy model where the cosmological constant is chosen so that the number of levels is two. In other words, the system is modelled as two interacting qubits. 

By treating them as two qubits, we successfully constructed the reduced density matrices both for 5D Sch-dS system and 4D Sch-dS system through entropic identification. To seek for the possible density matrix of the two-qubit system, we adopted the formulation developed in \cite{PhysRevA.93.062320}. The density matrix is firstly constructed as that of a product state of the two qubits. Then the density matrix undergoes non-local transformation, which introduces correlation between the two qubits so that it may be able to represent the Sch-dS system. Furthermore, by requiring that the density matrix of the combined system must be positive semi-definite, the non-local parameter $\theta_1$ must be constrained. The constraint, both from 4D and 5D system, then suggested a lower bound of the entropy of the combined system which is more stringent than the Araki-Lieb triangle inequality, shown in Fig. \ref{Fig:aCorr5D} and Fig. \ref{Fig:S5Dcompare} in 5D, and in Fig. \ref{Fig:aCorr4D} and Fig. \ref{Fig:S4Dcompare} in 4D. Corresponding to the quantum regime of correlation, this more stringent bound suggests that gravitational constraints can effect the quantum correlation in a way that it does not allow interacting qubits to be quantum-correlated as much as allowed by the Araki-Lieb triangle inequality.


\section*{Acknowledgement}

This research project has received funding support from the NSRF via the Program Management Unit for Human Resources \& Institutional Development, Research and Innovation [grant number B39G680009] and [grant number B39G680007].

\bibliography{refMISchdS}


\end{document}